\documentclass[12pt]{article}
\usepackage{graphicx}
\usepackage{setspace}
\usepackage[textwidth=16cm,textheight=22cm]{geometry}

\begin{document}
\newcommand{\Jpsi}{J/\psi}
\newcommand{\pT}{p_{T}}

\begin{center}
 {\Large \bf Overview of quarkonia and heavy flavour measurements by CMS } \\ 
\ \\
 {\large Prashant Shukla  for CMS collaboration} \\ 
\ \\
 {\it Nuclear Physics Division, Bhabha Atomic Research Center, 
 Mumbai, India }\\
  and \\
 {\it Homi Bhabha National Institute, Anushakti Nagar, Mumbai, India  } \\
\end{center}

\begin{abstract}
   This writeup summarizes recent CMS results on quarkonia measurements in pp, pPb and 
PbPb collisions at LHC. The excellent muon detection capability of CMS allows measurement 
of charmonia states at high transverse momentum ($p_T$) while the $\Upsilon$ states 
can be reconstructed starting at zero $\pT$.
  The absolute and relative yields of different charmonia and bottomonia states 
modified in PbPb collisions (over pp collisions) are described.
  The vertexing capability of CMS enables measurement of  
B meson energy loss via its decay to $\Jpsi$. 
  An overview of these measurements is given. How these measurements compare with 
other experiments at RHIC and LHC and have improved the understanding of 
heavy ion collisions has been discussed. 
\end{abstract}
\ \\
pacs:{\it 12.38.Mh, 24.85.+p, 25.75.-q} 
\ \\
keywords: {\it quark-gluon plasma, quarkonia, heavy flavour, charmonia, bottomonia, LHC} \\
\ \\
\ \\
Talk given at International Conference on Matter at Extreme Conditions : Then \& Now, \\
15-17 January 2014, Kolkata, India

\newpage

\section{Introduction}
  Heavy ion collisions at ultra-relativistic energies are performed to create and 
characterize quark gluon plasma (QGP), a phase of strongly interacting matter at an 
energy density where quarks and gluons are no longer bound within hadrons. 
  Quarkonia states ($\Jpsi$ and $\Upsilon$) have been one of the most popular tools 
since their suppression was proposed as a signal of QGP \cite{SATZ}.
  Quarkonia are produced early in the heavy ion collisions and if they evolve
through the deconfined medium their yields should be suppressed in comparison with those in pp. 
 The first such measurement was the 'anomalous' $\Jpsi$ suppression discovered in PbPb collisions 
at $\sqrt{s_{NN}}=17.3$ GeV at the SPS, which was considered as a hint of QGP formation. 
The RHIC measurements in AuAu at $\sqrt{s_{NN}}=200$ GeV \cite{PHENIXJPsi} showed almost the 
same suppression at a much higher energy contrary to the expectation \cite{Brambilla:2010cs}. 
 Such an observation was consistent with the scenario that at higher collision energy the 
expected greater suppression is compensated by regeneration of $\Jpsi$ by recombination of two 
independently produced charm quarks~\cite{Andronic_SH1}.

  After the LHC started PbPb collisions at $\sqrt{s_{NN}} = 2.76$ TeV, a wealth of
results have become available on quarkonia production \cite{QGP_Tc,Schukraft}.
  The suppression of quarkonia in PbPb collisions can quantify the colour screening 
properties  of strongly interacting matter \cite{SATZ} or alternatively the thermal 
gluon dissociation cross section of quarkonia \cite{BhanotPeskin,Xu}. 
  The statistical models \cite{Andronic_SH1} offer estimates of the regeneration of 
quarkonia from charm quark pairs. The inverse of the gluon dissociation process is also used to 
estimate regeneration \cite{Thews}.  
   There have been many recent calculations to explain the LHC results on quarkonia using a 
combination of above theoretical frameworks and models \cite{Rapp1,Rapp2}. 

   The CMS experiment with its muon detection capabilities has enabled several
measurements on quarkonia (both charmonia as well as bottomonia) via dimuon 
channel. The excellent mass resolution in dimuon channel allows
precise measurement of the three $\Upsilon$ states and their relative yields in pp, 
pPb as well as PbPb systems.
  Detailed measurements of $\Jpsi$ and $\psi(2S)$ have been made in different kinematic 
ranges. We give the results of these measurements and compare them with the other experiments 
at LHC and RHIC.
 The excellent vertexing capability of CMS enables measurement of B mesons 
via its decay to $\Jpsi$. The measurement of suppression of hadrons containing different quarks flavours 
can constrain various energy loss mechanisms \cite{ENLOSS}.

  The quarkonia yields in heavy ion collisions are also modified due to non-QGP effects such as
shadowing, an effect due to the change of the parton distribution functions inside the nucleus,
and dissociation due to hadronic or comover interactions \cite{Vogt}. To get a quantitative
idea about these effects, measurements in pPb collisions at 
$\sqrt{s_{NN}}=5.02$ TeV are performed some of them are discussed in this writeup.

\section{Charmonia measurements}

\begin{figure}
\begin{center}
\includegraphics[width=0.50\textwidth]{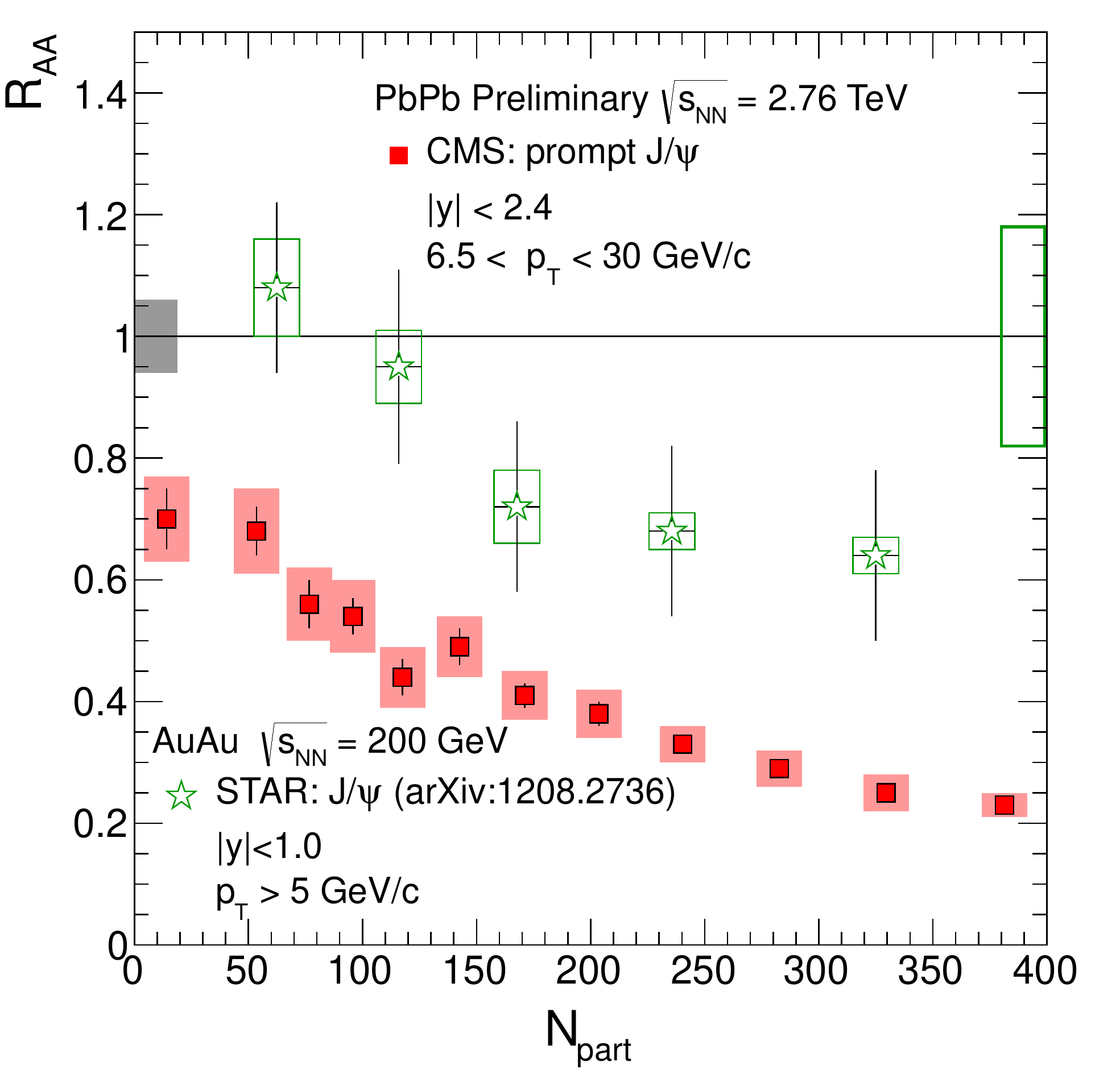}
\caption{The nuclear modification factor ($R_{\rm AA}$) of $\Jpsi$ in PbPb collisions 
at $\sqrt{s_{NN}} = 2.76$ TeV as a function of number of participants measured by CMS
experiment \cite{JCMS,CMSJPsi}. RHIC measurements are shown for comparison \cite{STARjpsi}.}
\label{fig:JPsiNpart}
\end{center}
\end{figure}

  The CMS experiment carries out $\Jpsi$ measurements at high transverse momentum 
($p_T>6.5$ GeV/$c$) and in the rapidity range $|y|\,\leq 2.4$.
  Figure~\ref{fig:JPsiNpart} shows the nuclear modification factor ($R_{\rm AA}$) of $\Jpsi$ in 
PbPb collisions at $\sqrt{s_{NN}} = 2.76$ TeV as a function of number of participants (centrality) 
measured by CMS \cite{JCMS,CMSJPsi}. 
 The $R_{\rm AA}$ of these high $p_T$ prompt $\Jpsi$ decreases with increasing 
centrality showing moderate suppression even in the most 
peripheral collisions. 
  On comparing with the STAR results \cite{STARjpsi} at RHIC, it follows that the 
suppression of (high $p_T$) $\Jpsi$ has increased with collision energy.
   The ALICE results on $\Jpsi$ correspond to a low $p_T$ range which 
have little or no centrality dependence except for the most peripheral collisions
\cite{ALICEJPsiNpart}. 

\begin{figure}
\begin{center}
\includegraphics[width=0.50\textwidth]{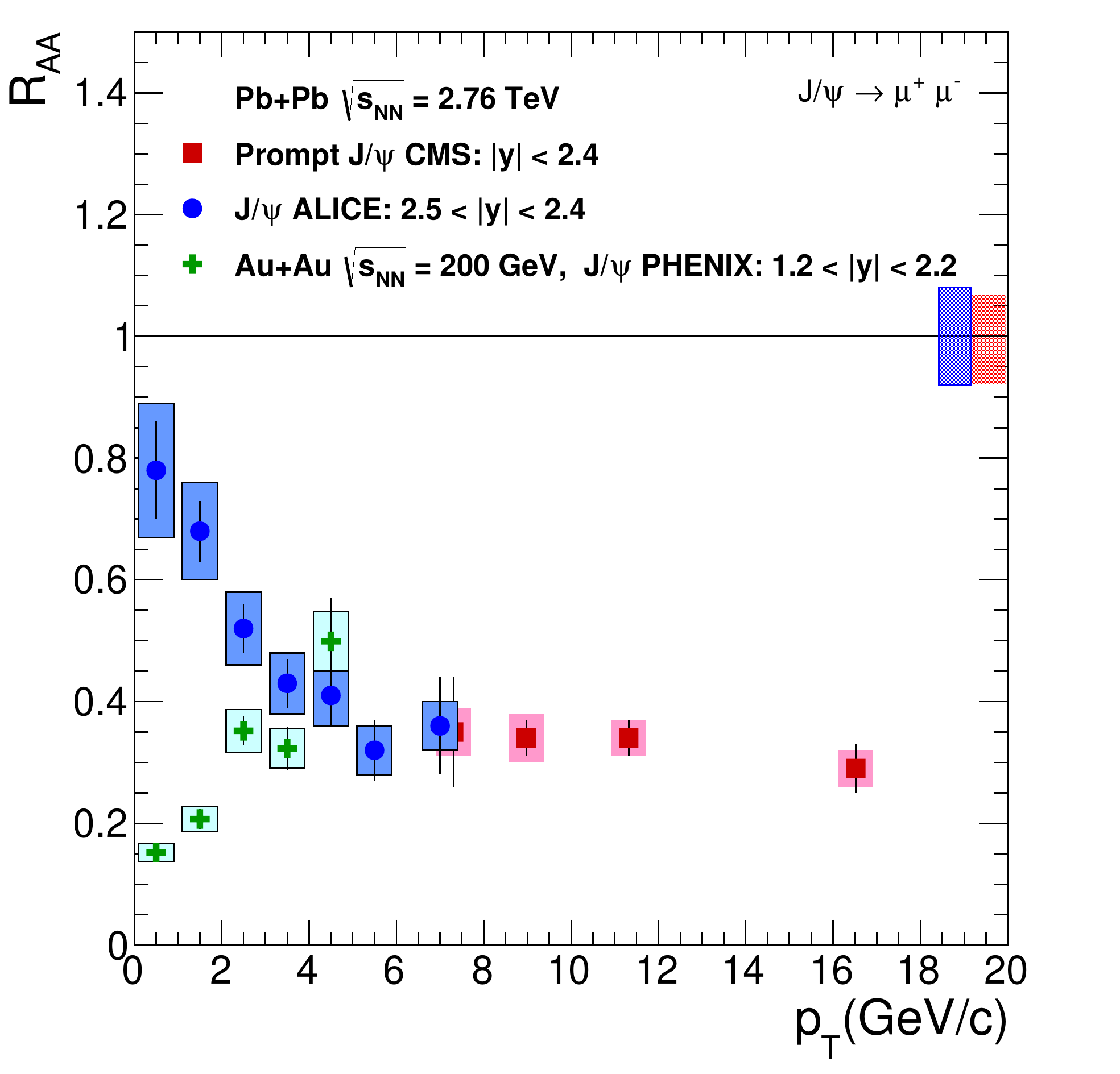}
\caption{ Nuclear modification factor ($R_{\rm AA}$) of $\Jpsi$ as a function of $p_T$ measured 
by CMS \cite{JCMS,CMSJPsi}, ALICE \cite{ALICEJPsi} and PHENIX \cite{PHENIXJPsi} experiments. }
\label{fig:JPsiPt}
\end{center}
\end{figure}

 Figure~\ref{fig:JPsiPt} shows $R_{\rm AA}$ of $\Jpsi$ in PbPb collisions 
at $\sqrt{s_{NN}} = 2.76$ TeV as a function of $p_T$ measured by CMS, ALICE 
and PHENIX experiments.  
The $R_{\rm AA}$ is found to be nearly independent of $p_T$ (above 6.5 GeV$/c$) showing 
that $\Jpsi$ remains suppressed even at very high $p_T$ upto 16 GeV/$c$
\cite{JCMS,CMSJPsi}.
 The ALICE $\Jpsi$ data  \cite{ALICEJPsi} shows that $R_{\rm AA}$ increases with 
decreasing $p_T$ below 
4 GeV/$c$. On comparing with the PHENIX forward rapidity measurement 
\cite{PHENIXJPsi}, it can be said that low p$_T$ $\Jpsi$ at LHC are
enhanced in comparison to RHIC.
  These observations suggest regeneration of $\Jpsi$ at low $p_T$ by 
recombination of independently produced charm pairs. Another hint of 
regeneration is given by CMS measurement of ratios of charmonia in PbPb and pp collisions. 

\begin{figure}
\begin{center}
\includegraphics[width=0.80\textwidth]{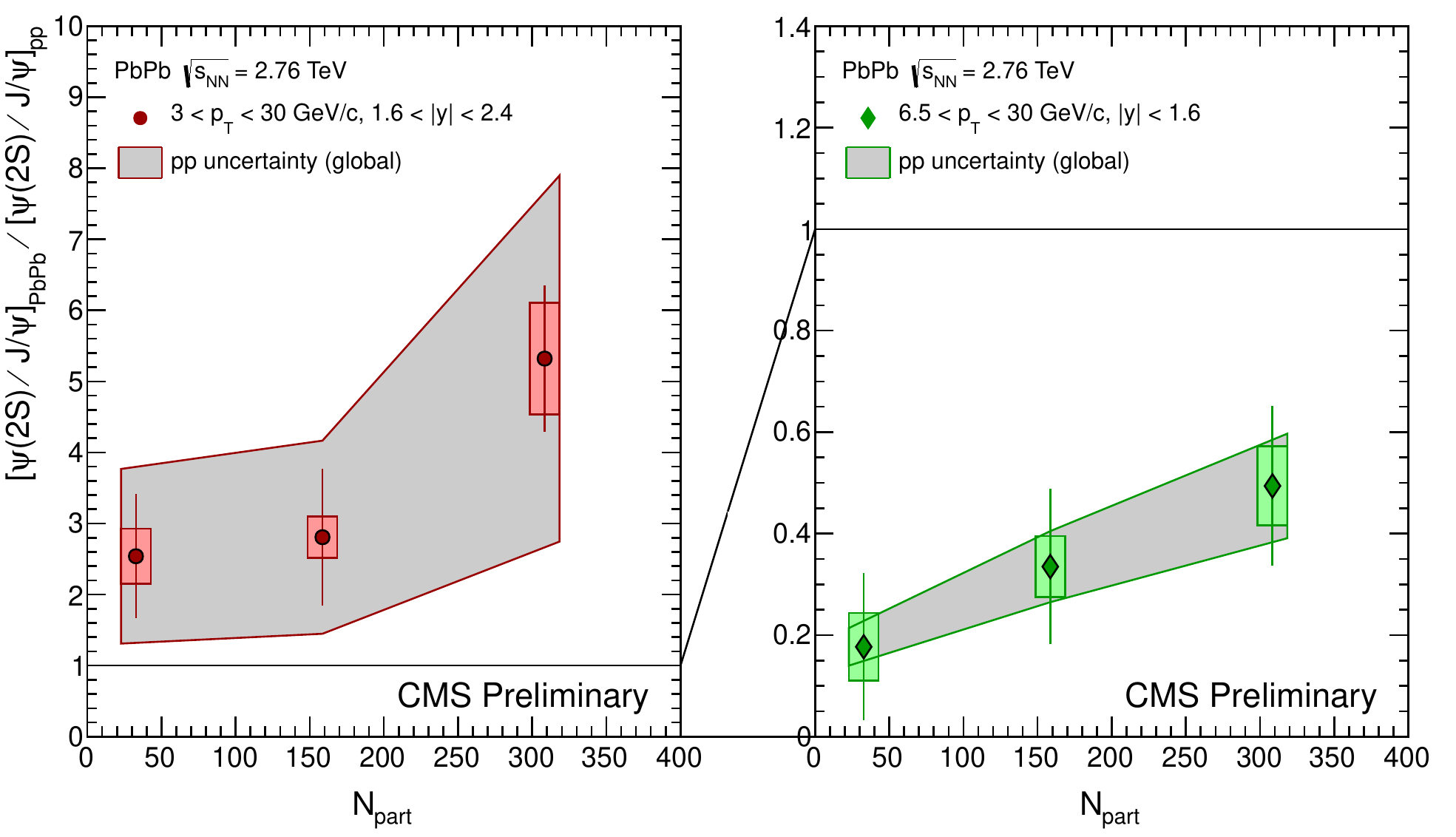}
\caption{Double ratio (ratio of ratios in PbPb to pp) of $\psi(2S)$ and $\Jpsi$ as a
function of centrality measured by CMS in two kinematic regions \cite{PSIP}.}
\label{fig:psiPrime}
\end{center}
\end{figure}

  Figure~\ref{fig:psiPrime} shows the double ratio of $\psi(2S)$ and $\Jpsi$ as a 
function of centrality measured by CMS in two kinematic regions \cite{PSIP}. 
The left plot is for 
low $p_T$ and forward rapidity region ($p_T > 3$ GeV/$c$ and $1.6<|y|< 2.4$) 
and the right is for high $p_T$ and central rapidity region
($p_T > 6.5$ GeV/$c$ and $|y|< 1.6$).
 Although there are large pp uncertainties, one can conclude that
at low $p_T$,  $\psi(2S)$ is less suppressed than $\Jpsi$ clearly for 
the most central collisions. Measurements with larger pp statistics will 
be able to confirm this conclusion.  

\section{Bottomonia measurements}

\begin{figure}
\begin{center}
\includegraphics[width=0.50\textwidth]{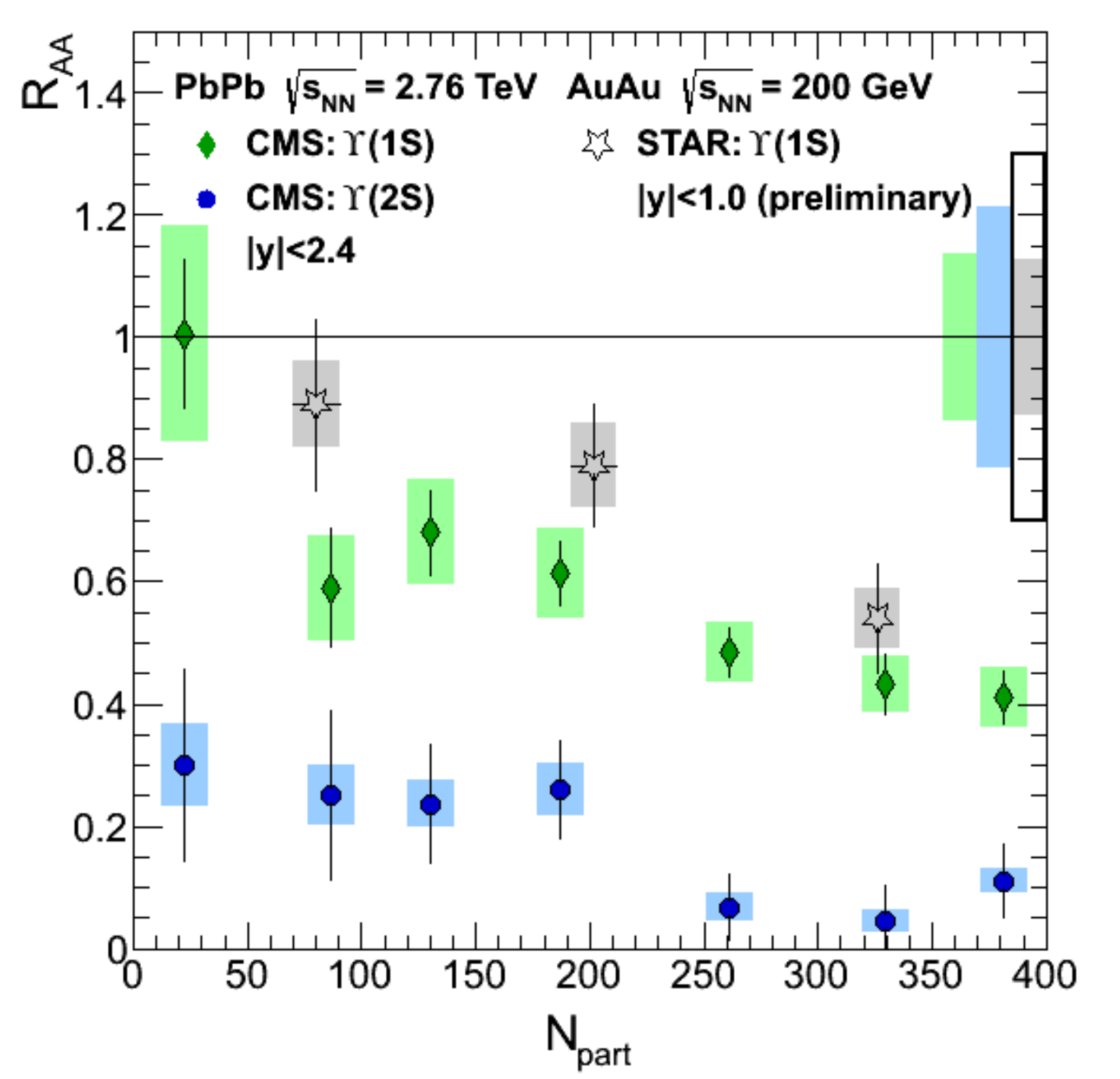}
\caption{ Nuclear modification factor ($R_{\rm AA}$) of $\Upsilon$(1S) and 
$\Upsilon$(1S) measured by CMS \cite{CMSUpsilon2}. RHIC measurements are plotted for 
comparison \cite{STARUP}.}
\label{fig:UpsilonNpart}
\end{center}
\end{figure}

  CMS measurements reveal that the higher $\Upsilon$ states are more suppressed 
relative to the ground state \cite{CMSUpsilon1,CMSUpsilon2} .
 This phenomenon is called sequential suppression where the bound states with 
smaller binding energies are more suppressed.
  Figure~\ref{fig:UpsilonNpart} shows the $R_{\rm AA}$ of $\Upsilon$(1S) and 
$\Upsilon$(2S) measured by CMS. The figure also shows STAR inclusive measurement 
of three $\Upsilon$ states \cite{STARUP}. The centrality integrated $R_{\rm AA}$ of 
$\Upsilon$(1S) state by CMS is $0.56 \pm 0.08 \pm 0.07$ as compared to  
$0.71 \pm 0.06 \pm 0.09$ by STAR which allows us 
to conclude that $\Upsilon$'s are more suppressed at higher collision energy.
 The new pp measurements made in 2013 will allow measurements of the $R_{\rm AA}$ of the $\Upsilon$ states as a function
of $p_T$ and rapidity.

\begin{figure}
\begin{center}
\includegraphics[width=0.50\textwidth]{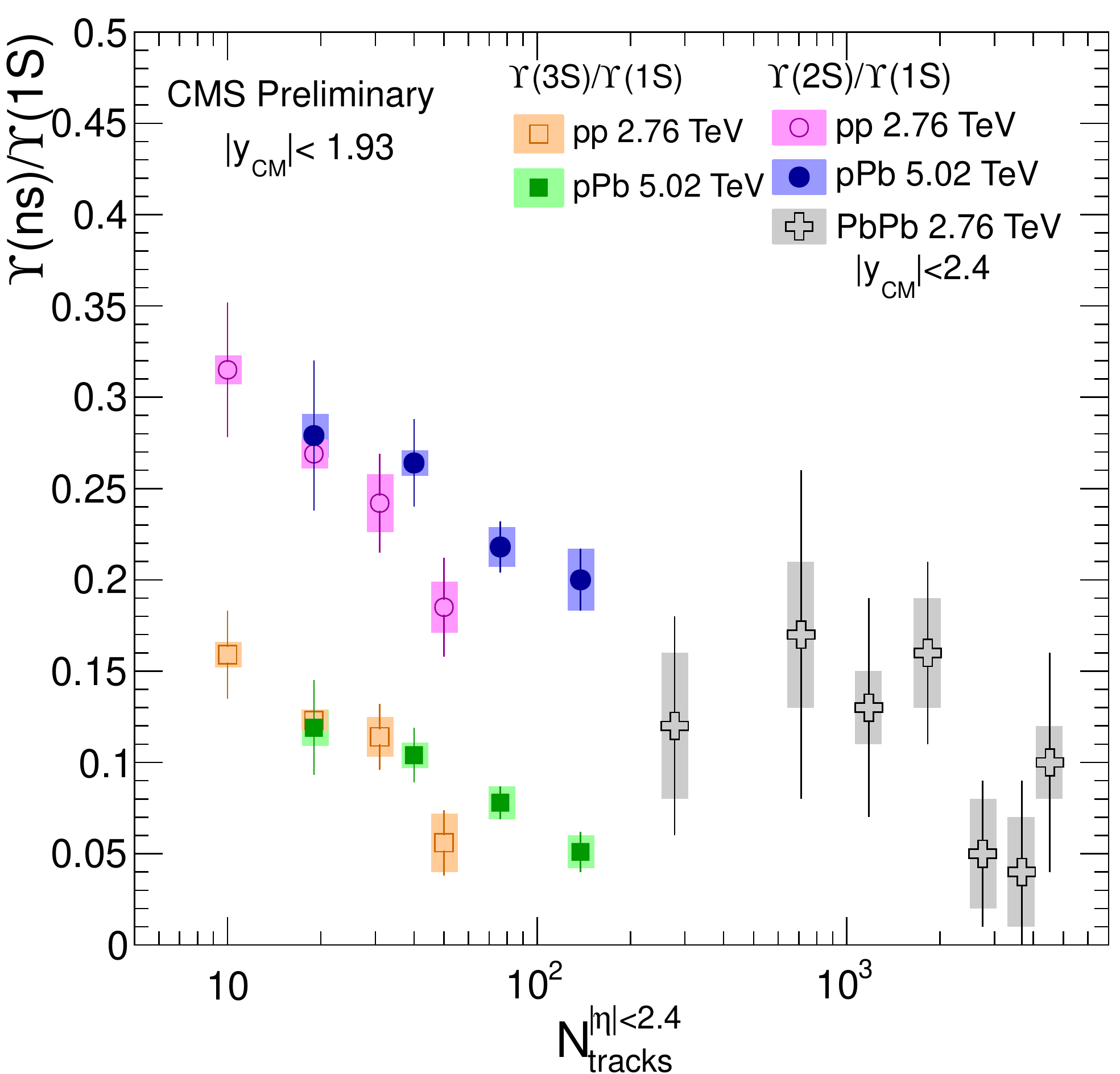}
\caption{The yield ratios $\Upsilon$(2S)/$\Upsilon$(1S) and $\Upsilon$(3S)/$\Upsilon$(1S)
as a function of number of track in event for different 
collision systems \cite{UPSPPB}.}
\label{fig:UpsilonpPb}
\end{center}
\end{figure}

 To study the effect of system size on the modification of quarkonia,
pPb collisions are performed at $\sqrt{s_{NN}}=5.02$ TeV with an integrated 
luminosity 5.4 (pb)$^{-1}$ \cite{UPSPPB}. 
 These measurements suggest the presence of final state effects in pPb collisions 
compared to pp collisions affecting ground state and excited states differently.
  Figure~\ref{fig:UpsilonpPb} shows 
the yield ratios $\Upsilon$(2S)/$\Upsilon$(1S) and $\Upsilon$(3S)/$\Upsilon$(1S)
as a function of number of tracks in the event for pp, pPb and PbPb collisions.
 The ratio seems to be constantly decreasing with increasing multiplicity. 
More PbPb data are needed to investigate the dependence in three systems and their 
possible relation.

\section{Heavy flavour measurements}

\begin{figure}
\begin{center}
\includegraphics[width=0.50\textwidth]{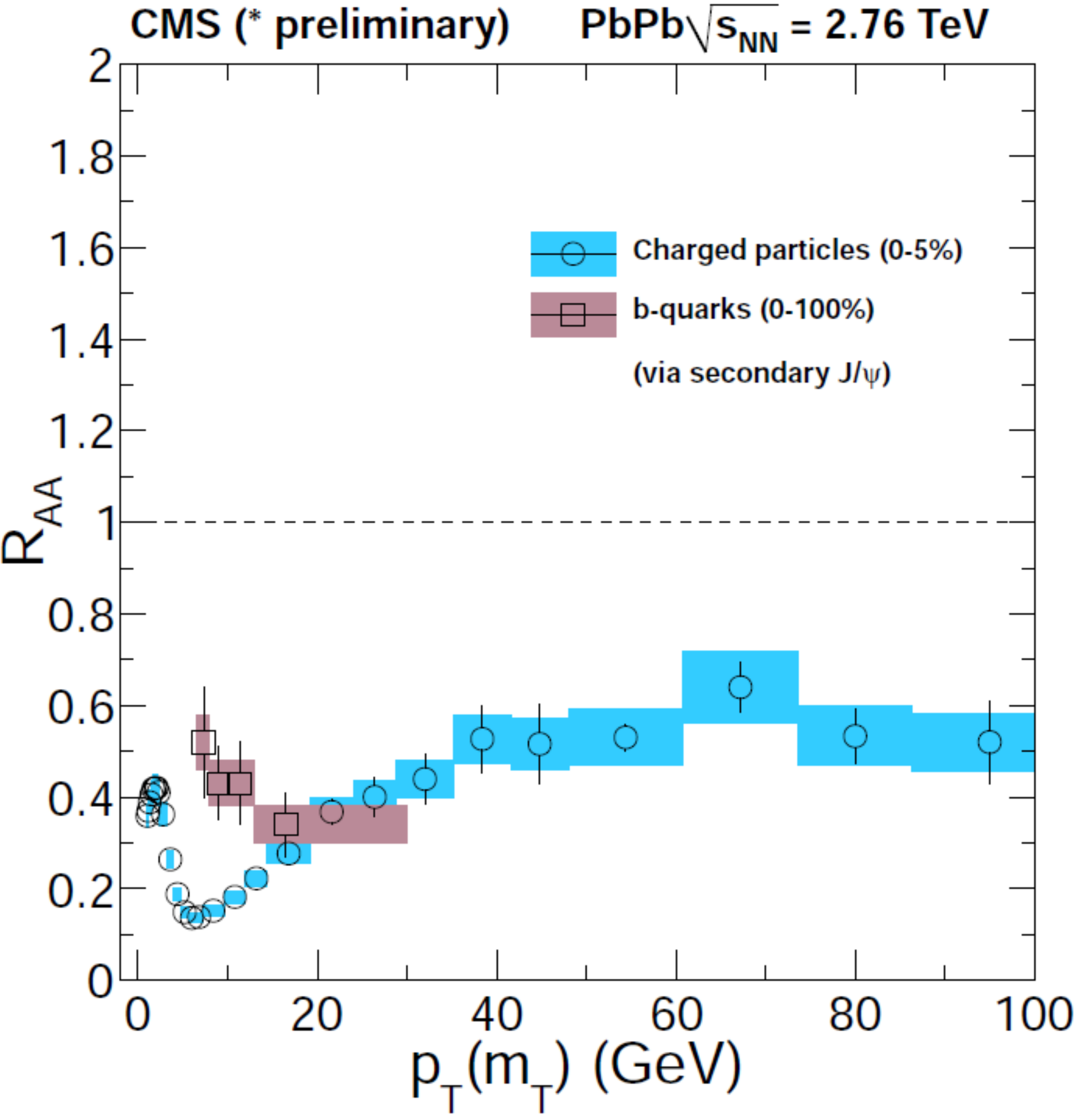}
\caption{Nuclear modification factor ($R_{\rm AA}$) of B mesons via secondary $\Jpsi$ compared to 
 $R_{\rm AA}$ of light charged hadrons \cite{JCMS,CMSJPsi}.}
\label{fig:Bmeson}
\end{center}
\end{figure}

  CMS offers B meson measurement via detecting secondary $\Jpsi$ coming from a displaced vertex. 
Figure~\ref{fig:Bmeson} shows the $R_{\rm AA}$ of B mesons via secondary $\Jpsi$ compared to 
$R_{\rm AA}$ of light hadrons \cite{JCMS,CMSJPsi}. We can conclude that at 
high $p_T > 10$ GeV/$c$ the suppression of B mesons and light hadrons are consistent, 
but at low $p_T$ B meson $R_{\rm AA}$ is larger as compared to light hadrons. Combining this
results with the ALICE measurements of D-meson \cite{ALICED} containing c-quarks it follows 
that at low $p_T$ there is mass hierarchy in the amount of suppression such that,
\ \\
 $R_{\rm AA}$ light hadrons $<$  $R_{\rm AA}$ D meson $<$ $R_{\rm AA}$ B meson. \\

\section{Summary}
  With the recent LHC measurements combined with RHIC measurements an overall 
understanding of quarkonia and heavy flavour production in heavy ion collisions is 
emerging. 
  One of the the most noticeable results is sequential suppression of $\Upsilon$ states 
observed first time in heavy ion collisions. The $\Upsilon$ suppression at LHC 
is more than that at RHIC showing that the matter at LHC has stronger colour screening. 
 The measurements of $\Upsilon$ states in pPb collisions suggest the presence of 
final effects in pPb collisions affecting ground state and excited states differently.

  High $p_T$ $\Jpsi$ is more suppressed at LHC as compared to RHIC.
 The enhancement of low $p_T$ $\Jpsi$ as compared to RHIC hints that that there
is substantial regeneration. The enhancement of ratio of yields of excited to ground
state charmonia at low $p_T$ also points in this direction.
 More statistics expected in PbPb collisions at 5 TeV, a better $p_T$ and rapidity dependence 
of quarkonia will certainly quantify the effects of colour screening and regeneration.

 The LHC hints mass hierarchy in suppression of hadrons 
below  $p_T \sim 8$ GeV/c. For $p_T > 10$ GeV/$c$, the suppression of light hadrons,
charm mesons and bottom mesons are  consistent. Better precision and larger 
$p_T$ reach will help quantifying the energy loss properties of the medium.

\end{document}